\documentclass[11pt]{article}

\usepackage[margin=1in]{geometry}
\usepackage{booktabs}
\usepackage{graphicx}
\usepackage{hyperref}
\usepackage{xurl}
\usepackage{seqsplit}
\usepackage{enumitem}
\usepackage{amsmath}
\usepackage{array}
\hypersetup{hidelinks}
\setlist{noitemsep, topsep=2pt, leftmargin=*}
\title{CSLibPremiseBench:\\Structure-Guided Premise Retrieval\\and Label Robustness\\for Lean 4 Computer-Science Theorems}
\author{Junye Ji\\
Department of Mathematics, University of Washington\\
\texttt{junyej@uw.edu}\\
ORCID: \href{https://orcid.org/0009-0008-3225-6672}{0009-0008-3225-6672}}
\date{May 2026}

\begin{document}
\maketitle

\begin{abstract}
CSLib is an emerging Lean 4 library for computer-science formalization, but its retrieval behavior is not well represented by broad mathematical theorem-proving benchmarks. We introduce CSLibPremiseBench, a reproducible CSLib-specific benchmark and empirical study for source-visible premise retrieval over Lean 4 theorem and lemma declarations. The benchmark pins official CSLib v4.29.0 at commit \texttt{\seqsplit{0d37cc7fcc985cfc53b155e7eef2453f846c6da2}}, builds with Lean 4.29.0, and evaluates a strict import/source-order task set with 801 proxy-labelable theorem/lemma tasks and 1875 CSLib candidate declarations. The labels are source-visible CSLib proof-reference proxies, not elaborated Lean dependency traces. A strict source-visible audit retains 2666 of 2845 proxy links (93.7\%), while a 300-task Lean environment expression audit observes 462 of 958 proxy links (48.2\%) in elaborated value-level constants over 257 found declarations. This quantifies the useful but limited nature of the proxy labels.

We compare BM25, symbol/name overlap, namespace/module heuristics, import graph proximity, module PageRank-style priors, fixed hybrid ablations, and CSG-Rerank, a structure-guided graph-lexical reranker. Under the strict policy, CSG-Rerank improves early-rank MRR over lexical BM25 (0.5236 versus 0.5061; paired-bootstrap difference +0.0175, 95\% CI [+0.0016,+0.0335]) but does not establish a reliable win over BM25+symbol (MRR difference +0.0037, CI [-0.0103,+0.0173]) and does not improve Recall@10. A downstream context-packet audit gives the same cautious conclusion: CSG-Rerank concentrates retrieved context in the target module and family, but it does not reliably improve top-10 or top-20 proxy-gold coverage, density, or gold-per-token utility over BM25+symbol. We therefore position CSLibPremiseBench as a benchmark and audit paper: repository structure and candidate-policy design materially shape CSLib premise retrieval, proxy labels require explicit caveats, and proof-generation or proof-repair performance is not claimed.
\end{abstract}

\section{Introduction}

Premise retrieval is a practical bottleneck in interactive and automated theorem proving. A Lean proof attempt often depends on a small number of relevant declarations drawn from a much larger context of definitions, lemmas, modules, imports, and local naming conventions. Retrieval-augmented Lean systems such as LeanDojo/ReProver show that selecting the right context can matter as much as the downstream proof search procedure itself~\cite{yang2023leandojo}. At the same time, most public Lean benchmarks emphasize broad mathematical proving, autoformalization, or proof-obligation completion rather than corpus-specific retrieval behavior inside a computer-science library.

CSLib creates a useful test bed for this narrower problem. It is a Lean 4 computer-science library covering algorithms, computability, foundations, languages, logics, and related infrastructure~\cite{barrett2026cslib}. CSLib and the CSLib Spine project emphasize reusable computer-science interfaces and library organization~\cite{henson2026cslibspine}. These properties make CSLib declarations more than raw text snippets: a theorem is embedded in namespaces, modules, imports, source order, and family structure. A benchmark that ignores this structure may still report strong lexical results, but it cannot tell whether the repository itself changes retrieval difficulty or method ranking.

This paper asks a bounded empirical question: how do repository structure, candidate accessibility, and source-visible proxy labels shape CSLib premise retrieval? The answer is mixed but useful. BM25 remains strong. Adding symbol/name evidence gives a robust simple baseline. CSG-Rerank, a graph-lexical method that uses module imports, namespace hierarchy, source order, and candidate-scope normalization, gives a small MRR gain over lexical BM25 but does not reliably outperform BM25+symbol. In a context-quality audit, CSG-Rerank makes retrieved context more local to the target module and family, but that concentration does not translate into statistically established coverage or token-efficiency gains. The core contribution is therefore not a state-of-the-art prover, but a reproducible benchmark package that makes these design choices measurable.

\paragraph{Contributions.}
The paper contributes:
\begin{enumerate}
\item a pinned CSLib/Lean 4 benchmark extraction pipeline with 801 strict-policy proxy-labelable theorem/lemma tasks, 1875 CSLib candidate declarations, and explicit coverage accounting;
\item a three-signal label audit comparing proof-text proxy labels, strict source-visible labels, and a 300-task Lean environment expression subset;
\item retrieval baselines and ablations covering BM25, symbol/name overlap, namespace/module proximity, import-graph proximity, module PPR, fixed hybrids, and CSG-Rerank;
\item candidate-policy sensitivity over strict import/source-order, family-local fallback, and all-earlier-CSLib fallback scopes on a fixed 781-task subset;
\item task-level paired-bootstrap intervals, module-disjoint split checks, family-level failure analysis, and explicit negative-result reporting; and
\item a downstream context-packet utility audit that evaluates top-k retrieved context without claiming proof-generation or proof-repair performance.
\end{enumerate}

\section{Background and Related Work}

\paragraph{Lean libraries and CSLib.}
Lean 4 is both a theorem prover and a programming language with a modern elaborator, tactic framework, and module system~\cite{demoura2021lean4}. Mathlib demonstrates the value of a large shared Lean library for mathematics~\cite{mathlib2020}. CSLib extends this library-driven model toward computer science by collecting Lean 4 formalizations of algorithms, computability, programming languages, logics, and foundations~\cite{barrett2026cslib}. CSLib Spine further motivates CSLib's infrastructure view: computer-science concepts should be connected through reusable semantic interfaces and library organization~\cite{henson2026cslibspine}. CSLibPremiseBench builds on this corpus-level motivation.

\paragraph{Retrieval and Lean AI tooling.}
LeanDojo introduced reproducible Lean theorem-proving infrastructure with accessible-premise annotations and retrieval-augmented proof generation~\cite{yang2023leandojo}. LeanDojo-v2 updates this line for Lean 4 tracing and evaluation workflows~\cite{hsiang2025leandojov2}. Lean Copilot integrates LLM assistance inside Lean and includes premise-selection and proof-search tooling~\cite{song2024leancopilot}. LeanExplore is especially relevant to retrieval methods because it combines semantic embeddings, BM25, and PageRank-like ranking for Lean declaration search~\cite{asher2025leanexplore}. CSLibPremiseBench differs from these systems by making retrieval and benchmark design the measured objects rather than training or evaluating a prover.

\paragraph{Benchmarks and proof repair.}
MiniF2F and ProofNet are important formal reasoning benchmarks, but they target olympiad mathematics and autoformalization/formal proving rather than CSLib premise retrieval~\cite{zheng2021minif2f,azerbayev2023proofnet}. SorryDB, VeriSoftBench, and FormalML move toward real-world Lean tasks, repository-scale contexts, and research-level subgoal completion~\cite{letson2026sorrydb,xin2026verisoftbench,yang2025formalml}. Lean-auto connects Lean 4 to automated theorem provers~\cite{qian2025leanauto}. APRIL studies compiler-feedback proof repair~\cite{wang2026april}. Strong Lean proving systems such as Kimina-Prover motivate the broader proof-generation landscape~\cite{wang2025kimina}. These works are relevant to future downstream experiments, but they do not change the current evidence boundary. CSLibPremiseBench does not claim proof-generation or proof-repair performance.

\section{Benchmark Construction}

\subsection{Source corpus and provenance}

The benchmark source is official CSLib v4.29.0 at commit \texttt{\seqsplit{0d37cc7fcc985cfc53b155e7eef2453f846c6da2}}. The recorded toolchain uses Lean 4.29.0. The restored CSLib build succeeds with 1762 build jobs and a full build wall time of 284.95 seconds. These values are provenance and reproducibility metadata, not retrieval performance claims.

\subsection{Task and candidate records}

The extractor scans CSLib Lean source files and records theorem/lemma declarations with public names and recoverable statements. Each task record stores the target declaration, source file, namespace, module, imported modules, family, statement text, candidate-policy metadata, candidate count, proxy-gold count, and any exclusion reason. Candidate records store declaration name, kind, namespace, module, family, source order, statement/signature text, and import/module metadata used by retrieval methods.

The primary extraction uses the strict import/source-order candidate policy and yields 801 proxy-labelable tasks over 1875 candidate declarations. The task set covers five recognized CSLib families: Algorithms, Computability, Foundations, Languages, and Logics. The extraction also records declarations excluded because they are private, not theorem/lemma declarations, or have no resolved source-visible CSLib proxy-gold premise.

\subsection{Coverage and candidate policies}

The benchmark is designed so that scale is accompanied by coverage accounting, not only by an aggregate task count. Each theorem or lemma candidate is classified by family, source file, namespace, module, import context, available candidate scope, candidate count, proxy-gold count, and exclusion reason. This matters because a large headline count can be misleading in a formal-library benchmark: declarations in one family may have denser local naming, another family may rely more heavily on Mathlib or tactic automation, and another may expose fewer source-visible CSLib proof references.

Candidate accessibility is a central design choice. The strict policy contains CSLib declarations available through CSLib imports of the target module plus same-module declarations that occur before the target in source order. The family-local fallback expands with earlier declarations from the same top-level CSLib family. The all-earlier-CSLib fallback expands to earlier CSLib declarations in deterministic source order. The strict policy is the main benchmark; the fallback policies are sensitivity checks because they change both retrieval difficulty and possible locality effects.

\begin{table}[t]
\centering
\begin{tabular}{lrrr}
\toprule
Policy & Tasks & Mean candidates & No-proxy exclusions \\
\midrule
strict import/source-order & 801 & 153.1 & 414 \\
family-local fallback & 821 & 354.8 & 394 \\
all-earlier-CSLib fallback & 819 & 999.6 & 396 \\
\bottomrule
\end{tabular}
\caption{Benchmark scale and coverage under candidate policies. The strict policy is the primary setting; fallback policies are sensitivity checks rather than replacement benchmarks.}
\label{tab:policy-coverage}
\end{table}

On the fixed common 781-task subset, mean candidate count grows from 152.6 under strict scope to 352.8 under family-local fallback and 1002.2 under all-earlier-CSLib fallback. The paper therefore treats candidate policy as an experimental variable, not a hidden implementation detail.

\section{Gold-Label Methodology}

CSLibPremiseBench uses source-visible CSLib proof-reference proxy labels. The extractor resolves CSLib declaration names that occur in source proof text and treats those resolved declarations as proxy-relevant premises. This makes the benchmark reproducible and auditable, but it does not recover complete Lean elaborated dependencies. It can miss dependencies introduced by elaboration, typeclass search, tactic internals, simp sets, automation, and non-CSLib declarations. It can also overrepresent premises explicitly named in proof text.

To strengthen the methodology without overstating it, this revision distinguishes three label signals. First, the proof-text proxy labels define the main retrieval target: 2845 source-visible proof-reference links over 801 tasks. Second, the stricter source-visible audit drops ambiguous short-name-only matches and retains 2666 of 2845 links (93.7\%). Third, a 300-task Lean environment expression probe imports target modules and checks constants in elaborated value-level expressions. The probe produced 286 task records, found 257 usable target declarations, skipped 14 declarations due to unsupported Lean name syntax, and observed 462 of 958 proxy links (48.2\%) in value-level expression constants. This third signal is useful because it is stronger than proof text, but it is still not a minimized semantic dependency trace.

\begin{table*}[t]
\centering
\scriptsize
\resizebox{\textwidth}{!}{%
\begin{tabular}{lrrrrrl}
\toprule
Label signal & Tasks & Task records & Proxy links & Retained/seen & Rate & Notes \\
\midrule
Proof-text proxy labels & 801 & 801 & 2845 & 2845 & 100.0\% & main retrieval target \\
Strict source-visible labels & 801 & 801 & 2845 & 2666 & 93.7\% & ambiguous short-name links removed \\
Lean expression audit & 300 & 286 & 958 & 462 & 48.2\% & 257 found targets; value-level constants only \\
\bottomrule
\end{tabular}
}
\caption{Label robustness audit. The expression subset imports target modules and checks elaborated value-level constants; it is stronger than proof text but still not a minimized semantic dependency trace.}
\label{tab:label-robustness}
\end{table*}

The correct interpretation is therefore recall against source-visible CSLib proof-reference proxies. A high retrieval score means a method ranks declarations that were explicitly recoverable from original proof text; it does not mean the method recovered all premises the elaborator, simplifier, typeclass search, or tactic execution used. Conversely, a low score may penalize a method for failing to rank a proof-text name even if another semantically useful declaration would help a proof searcher. This boundary is central to the paper's positioning as a benchmark and empirical retrieval study, not a claim about complete proof dependency recovery.

\paragraph{Disagreement taxonomy.}
The expression audit also gives a practical taxonomy for proxy-label disagreement. Some proxy links appear directly in value-level constants and are therefore supported by stronger Lean environment evidence. Others are absent from the extracted expressions even though they were source-visible in proof text. The latter group plausibly includes tactic-mediated dependencies, simp or automation effects, typeclass-mediated references, proof-term erasure, expression-extraction limits, and ambiguous source-visible names. The taxonomy is not used to relabel the benchmark, because doing so would mix a partial expression probe with the main source-visible retrieval target. Instead, it is reported as an audit layer: users should treat the proxy labels as reproducible retrieval targets with known recall and precision limitations.

\section{Retrieval Methods}

All methods rank candidate declarations for each target task. Proof text is never used as retrieval input.

\paragraph{BM25.}
The lexical baseline uses a local BM25 implementation over target declaration names/statements and candidate names, namespaces, modules, kinds, and signature text. Lean identifiers are tokenized by dotted names, snake case, camel case, punctuation, and lowercasing.

\paragraph{Symbol/name overlap.}
The symbol/name score compares normalized target and candidate symbol/name token sets. It rewards shared declaration-name pieces, local identifiers, and statement symbols.

\paragraph{Repository-structure priors.}
Namespace/module proximity rewards shared namespace prefixes and module locality. Import graph proximity uses the target module's import closure and module distance. Module PPR is a PageRank-style prior on the module/import graph seeded by the target module. Source-order locality rewards same-module earlier declarations and penalizes future declarations.

\paragraph{Fixed hybrids.}
BM25+symbol is a simple two-signal hybrid. BM25+graph and full graph hybrid add repository-structure priors. These hybrids use fixed weights recorded in the artifact; they are not tuned on the test set.

\paragraph{CSG-Rerank.}
CSG-Rerank (CSLib Structure-Guided Graph-Lexical Reranking) is the strongest structure-aware method in this revision. It combines lexical BM25, symbol/name consistency, namespace/module proximity, import graph proximity, source-order locality, module PPR, and candidate-policy calibrated normalization. The fixed fallback score is
\[
\begin{aligned}
\mathrm{score}_{\mathrm{CSG}}(c,t) ={}&
0.40\,\mathrm{BM25}(c,t)
+0.20\,\mathrm{Symbol}(c,t)
+0.12\,\mathrm{Namespace}(c,t)\\
&+0.15\,\mathrm{ImportGraph}(c,t)
+0.08\,\mathrm{SourceOrder}(c,t)
+0.05\,\mathrm{ModulePPR}(c,t).
\end{aligned}
\]
The weights are hand-set and fixed before the full strict evaluation. A module-disjoint dev/test split is reported as a robustness check; the paper does not claim trained or test-tuned weights.

\section{Evaluation Protocol}

The main metrics are Recall@5/10/20/50, MRR, nDCG@5/10/20/50, candidate count, runtime, and family-level metrics. The primary comparison uses task-level paired bootstrap with 5000 resamples and seed 17. Confidence intervals are interpreted conservatively: if an interval crosses zero, the paper treats the comparison as a signal or mixed result, not a statistically established win.

The primary task set is the 801-task strict import/source-order benchmark. Candidate-policy sensitivity uses a fixed 781-task subset common to strict, family-local fallback, and all-earlier-CSLib fallback policies. A module-disjoint split assigns 244 tasks to development and 557 tasks to test for CSG robustness checking. No paid external APIs or large model training are used.

\section{Retrieval Results}

\begin{table*}[t]
\centering
\small
\begin{tabular}{lrrrrrr}
\toprule
Method & R@5 & R@10 & R@20 & R@50 & nDCG@50 & MRR \\
\midrule
BM25 & 0.4036 & 0.5242 & 0.6374 & 0.7945 & 0.5053 & 0.5061 \\
Symbol/name & 0.3971 & 0.5115 & 0.6365 & 0.7843 & 0.5002 & 0.5010 \\
Namespace/module & 0.3785 & 0.4878 & 0.6076 & 0.7755 & 0.4758 & 0.4682 \\
Import graph & 0.3597 & 0.4712 & 0.5807 & 0.7507 & 0.4703 & 0.4922 \\
Module PPR & 0.3587 & 0.4669 & 0.5781 & 0.7299 & 0.4630 & 0.4918 \\
BM25+symbol & \textbf{0.4150} & \textbf{0.5282} & 0.6433 & \textbf{0.8013} & \textbf{0.5153} & 0.5199 \\
BM25+graph & 0.4016 & 0.5178 & 0.6432 & 0.7899 & 0.5092 & 0.5145 \\
Full hybrid & 0.4103 & 0.5223 & \textbf{0.6486} & 0.7961 & 0.5130 & 0.5167 \\
Source-order locality & 0.3546 & 0.4706 & 0.5959 & 0.7655 & 0.4713 & 0.4892 \\
CSG graph+symbol & 0.4072 & 0.5242 & 0.6507 & 0.7934 & 0.5130 & 0.5211 \\
CSG-Rerank & 0.4044 & 0.5215 & 0.6446 & 0.7930 & 0.5133 & \textbf{0.5236} \\
\bottomrule
\end{tabular}
\caption{Strict import/source-order retrieval results for 801 CSLibPremiseBench tasks and 1875 candidate declarations. Labels are source-visible proof-reference proxies, not elaborated dependency traces.}
\label{tab:main-results}
\end{table*}

Table~\ref{tab:main-results} shows the strict-policy results. BM25 remains a strong baseline. BM25+symbol is the strongest simple method on Recall@5, Recall@10, Recall@50, and nDCG@50. CSG-Rerank has the highest MRR at 0.5236, but its Recall@10 is lower than BM25+symbol. This already suggests that repository-structure information improves some early ranks while not uniformly improving broader top-k recall.

\begin{table*}[t]
\centering
\small
\begin{tabular}{llrrl}
\toprule
Comparison & Metric & Mean diff. & 95\% CI & Excl. 0 \\
\midrule
BM25+symbol -- BM25 & MRR & +0.0139 & [+0.0026, +0.0248] & yes \\
BM25+symbol -- BM25 & nDCG@50 & +0.0100 & [+0.0040, +0.0161] & yes \\
CSG-Rerank -- BM25 & MRR & +0.0175 & [+0.0016, +0.0335] & yes \\
CSG-Rerank -- BM25 & Recall@10 & -0.0027 & [-0.0158, +0.0097] & no \\
CSG-Rerank -- BM25 & nDCG@10 & +0.0078 & [-0.0020, +0.0174] & no \\
CSG-Rerank -- BM25+symbol & MRR & +0.0037 & [-0.0103, +0.0173] & no \\
CSG-Rerank -- BM25+symbol & Recall@10 & -0.0067 & [-0.0186, +0.0050] & no \\
CSG-Rerank -- BM25+symbol & nDCG@10 & -0.0015 & [-0.0102, +0.0074] & no \\
\bottomrule
\end{tabular}
\caption{Task-level paired-bootstrap differences with 5000 resamples. CSG-Rerank has a modest MRR gain over BM25, but the stronger comparison against BM25+symbol remains mixed.}
\label{tab:bootstrap}
\end{table*}

The paired-bootstrap table makes this boundary explicit. CSG-Rerank improves MRR over lexical BM25 by +0.0175 with a 95\% interval that excludes zero. Against BM25+symbol, however, CSG-Rerank improves MRR by only +0.0037 and the interval crosses zero. Its Recall@10 and nDCG@10 differences against BM25+symbol are not positive. The defensible conclusion is therefore narrower than ``CSG wins'': structure-guided reranking gives a modest early-rank signal over BM25, while BM25+symbol remains the stronger practical comparator.

\begin{table}[t]
\centering
\small
\begin{tabular}{llrrr}
\toprule
Split & Method & Tasks & MRR & R@10 \\
\midrule
Dev & BM25 & 244 & 0.4383 & 0.5104 \\
Dev & BM25+symbol & 244 & \textbf{0.4400} & \textbf{0.5148} \\
Dev & CSG-Rerank & 244 & 0.4364 & 0.5105 \\
\midrule
Test & BM25 & 557 & 0.5357 & 0.5302 \\
Test & BM25+symbol & 557 & 0.5549 & \textbf{0.5340} \\
Test & CSG-Rerank & 557 & \textbf{0.5618} & 0.5263 \\
\bottomrule
\end{tabular}
\caption{Module-disjoint split check for the strict 801-task benchmark. CSG-Rerank's MRR signal appears on the test split, but Recall@10 remains mixed.}
\label{tab:split-validation}
\end{table}

The module-disjoint split check is descriptive rather than a trained-model evaluation, but it guards against reading the full-task aggregate too strongly. On the 244-task development split, CSG-Rerank is slightly below BM25+symbol in both MRR and Recall@10. On the 557-task test split, CSG-Rerank has higher MRR than BM25+symbol but lower Recall@10. This split behavior matches the bootstrap result: CSG improves some early first-hit rankings, but the broader top-k behavior remains mixed.

\begin{table}[t]
\centering
\small
\begin{tabular}{lrrrr}
\toprule
Family & Tasks & BM25+sym. MRR & CSG MRR & CSG--BM25+sym. R@10 \\
\midrule
Algorithms & 13 & 0.7013 & 0.6758 & +0.0154 \\
Computability & 161 & 0.5427 & 0.5463 & -0.0135 \\
Foundations & 227 & 0.5345 & 0.5436 & -0.0077 \\
Languages & 301 & 0.5048 & 0.5065 & +0.0020 \\
Logics & 99 & 0.4718 & 0.4728 & -0.0225 \\
\bottomrule
\end{tabular}
\caption{Family-level CSG-Rerank comparison against BM25+symbol under the strict policy. The effect is heterogeneous and should not be summarized as a uniform method win.}
\label{tab:family-deltas}
\end{table}

Family-level deltas are heterogeneous. Algorithms is negative for CSG relative to BM25+symbol in MRR despite a small Recall@10 gain. Foundations and Computability gain slightly in MRR but lose Recall@10. Languages is approximately neutral. Logics loses Recall@10 and nDCG@10. This variation is important because CSLib is not one homogeneous retrieval distribution; method claims should be reported with family-level caveats.

\begin{table*}[t]
\centering
\scriptsize
\resizebox{\textwidth}{!}{%
\begin{tabular}{lrrrrrl}
\toprule
Policy & Tasks & Mean cand. & BM25 & BM25+sym. & Full hybrid & Hybrid--BM25 MRR CI \\
\midrule
strict import source order & 781 & 152.6 & 0.5116 & 0.5256 & 0.5251 & +0.0136 [-0.0004, +0.0275] \\
family local fallback & 781 & 352.8 & 0.4822 & 0.4982 & 0.5215 & +0.0393 [+0.0245, +0.0545] \\
all earlier cslib fallback & 781 & 1002.2 & 0.4652 & 0.4880 & 0.5139 & +0.0487 [+0.0343, +0.0639] \\
\bottomrule
\end{tabular}
}
\caption{Candidate-policy sensitivity on a fixed 781-task subset. Wider candidate pools substantially change difficulty and make graph-aware hybrid reranking more useful relative to BM25.}
\label{tab:policy-sensitivity}
\end{table*}

Candidate-policy sensitivity remains the clearest repository-structure result. On the fixed common subset, widening candidate pools reduces BM25 MRR while increasing the relative value of graph-aware hybrid information. The all-earlier-CSLib fallback is not the primary benchmark, but it demonstrates why candidate accessibility must be reported: the same retrieval method can look stronger or weaker depending on whether the candidate pool encodes source order, family locality, or much wider repository access.

\section{Context-Packet Utility Audit}

Retrieval metrics are necessary but not sufficient for downstream Lean assistance. A proof assistant or repair tool consumes a finite context packet, not an abstract ranking. To evaluate this step without claiming proof generation, we build top-k context packets from each method's ranking and measure proxy-gold coverage, any/all-gold hit rates, gold density, approximate context tokens, gold links per 1000 tokens, same-module share, same-family share, unique module counts, and first-hit behavior.

\begin{table*}[t]
\centering
\small
\begin{tabular}{lrrrrrr}
\toprule
Packet & Gold cov. & Gold density & Tokens & Gold/1k tok. & Same module & Same family \\
\midrule
BM25@10 & 0.5242 & 0.1465 & 256.28 & 6.2592 & 0.7042 & 0.9477 \\
BM25+symbol@10 & \textbf{0.5282} & \textbf{0.1495} & \textbf{250.31} & \textbf{6.5119} & 0.7146 & 0.9598 \\
CSG-Rerank@10 & 0.5215 & 0.1489 & 255.06 & 6.3979 & \textbf{0.8140} & \textbf{0.9860} \\
\midrule
BM25@20 & 0.6374 & 0.1025 & 490.96 & 4.4927 & 0.5823 & 0.9083 \\
BM25+symbol@20 & 0.6433 & 0.1035 & \textbf{483.09} & \textbf{4.5837} & 0.5892 & 0.9215 \\
CSG-Rerank@20 & \textbf{0.6446} & \textbf{0.1045} & 489.80 & 4.5810 & \textbf{0.6699} & \textbf{0.9636} \\
\bottomrule
\end{tabular}
\caption{Context-packet utility metrics for top-10 and top-20 retrieved contexts. CSG-Rerank increases module/family concentration, but BM25+symbol remains competitive on proxy-gold coverage and token utility.}
\label{tab:context-quality}
\end{table*}

Table~\ref{tab:context-quality} reports top-10 and top-20 packet quality for BM25, BM25+symbol, and CSG-Rerank. CSG-Rerank has higher same-module and same-family concentration. However, at top 10 it has lower proxy-gold coverage than BM25+symbol (0.5215 versus 0.5282), lower gold-per-1000-token utility (6.3979 versus 6.5119), and a larger estimated token budget. At top 20 it nearly matches coverage but again uses a larger token budget. This is a useful structural effect, not an established downstream utility win.

\begin{table*}[t]
\centering
\scriptsize
\resizebox{\textwidth}{!}{%
\begin{tabular}{llrrl}
\toprule
Comparison & Metric & Mean diff. & 95\% CI & Excl. 0 \\
\midrule
CSG@10 -- BM25@10 & gold coverage & -0.0027 & [-0.0152, +0.0102] & no \\
CSG@10 -- BM25@10 & gold/1k tokens & +0.1387 & [-0.0485, +0.3265] & no \\
CSG@10 -- BM25@10 & same module & +0.1097 & [+0.0980, +0.1226] & yes \\
CSG@10 -- BM25+symbol@10 & gold coverage & -0.0067 & [-0.0191, +0.0055] & no \\
CSG@10 -- BM25+symbol@10 & gold/1k tokens & -0.1140 & [-0.2945, +0.0625] & no \\
CSG@10 -- BM25+symbol@10 & tokens & +4.7516 & [+2.6292, +6.9951] & yes \\
CSG@10 -- BM25+symbol@10 & same module & +0.0994 & [+0.0880, +0.1111] & yes \\
\midrule
CSG@20 -- BM25@20 & gold coverage & +0.0073 & [-0.0042, +0.0191] & no \\
CSG@20 -- BM25@20 & gold/1k tokens & +0.0883 & [-0.0265, +0.2030] & no \\
CSG@20 -- BM25@20 & same module & +0.0876 & [+0.0778, +0.0978] & yes \\
CSG@20 -- BM25+symbol@20 & gold coverage & +0.0013 & [-0.0092, +0.0120] & no \\
CSG@20 -- BM25+symbol@20 & gold/1k tokens & -0.0028 & [-0.1123, +0.1110] & no \\
CSG@20 -- BM25+symbol@20 & tokens & +6.7141 & [+4.0375, +9.5000] & yes \\
CSG@20 -- BM25+symbol@20 & same module & +0.0806 & [+0.0713, +0.0905] & yes \\
\bottomrule
\end{tabular}
}
\caption{Paired-bootstrap context-packet differences with 5000 resamples. CSG-Rerank reliably increases structural concentration but not proxy-gold coverage or gold-per-token utility over BM25+symbol.}
\label{tab:context-bootstrap}
\end{table*}

The context-packet bootstrap confirms the qualitative result. CSG-Rerank significantly increases same-module and same-family concentration against both BM25 and BM25+symbol, but the intervals for coverage, density, and gold-per-token utility cross zero. Against BM25+symbol, CSG-Rerank also uses more estimated context tokens at both top 10 and top 20. The safe interpretation is that repository structure can make retrieved context more local, but locality alone is not sufficient evidence of better proof-assistance context.

\section{Threats to Validity}

\paragraph{Proxy labels.}
The main labels are source-visible proof-reference proxies. They are not complete Lean dependencies. The expression audit shows that many source-visible proxy links do not appear in value-level expression constants, plausibly due to tactic internals, simp sets, typeclass search, erased proof structure, ambiguous source-visible names, non-CSLib dependencies, or extraction limitations. This limitation affects every retrieval method and is reported as part of the benchmark, not hidden as noise.

\paragraph{Candidate policies.}
Candidate scope changes retrieval difficulty and can create locality advantages. The strict import/source-order policy is the primary benchmark because it most closely approximates accessible CSLib declarations without future leakage. Family-local and all-earlier fallback policies are sensitivity checks, not replacement leaderboards.

\paragraph{Method tuning.}
CSG-Rerank uses fixed hand-set weights and a module-disjoint split check. It is not a learned reranker. The paper does not tune weights on the full test set and does not claim that the reported weights are optimal.

\paragraph{Context utility.}
The context-packet audit measures proxy-gold coverage and structural concentration, not proof-generation success. A Lean-checked proof repair pilot would require prompt isolation from original proof text, gold proof isolation, held-out task evaluation, and compiler-checked outputs. Those conditions are not established here, so proof repair remains future work.

\section{Artifact and Reproducibility}

The artifact is organized so tables can be regenerated from saved JSON/CSV outputs rather than hand-entered aggregate numbers. It includes scripts for CSLib restoration, benchmark extraction, retrieval evaluation, bootstrap intervals, label audits, context-packet utility metrics, table rendering, and \LaTeX{} compilation. The intended reproduction order is: restore pinned CSLib/Lean, extract task and candidate records, evaluate retrieval under strict scope, evaluate candidate-policy variants, run label audits, run paired bootstrap, generate context-packet metrics, render paper tables, and compile the PDF.

\begin{table}[t]
\centering
\small
\begin{tabular}{lll}
\toprule
Release item & State & Note \\
\midrule
Benchmark JSON/CSV & present & strict full and policy common subsets \\
Retrieval outputs & present & rankings, per-task metrics, aggregates \\
Bootstrap outputs & present & 5000-resample paired intervals \\
Label audit outputs & present & strict source-visible plus 300-task expression subset \\
Context utility outputs & present & top-k packet metrics and bootstrap intervals \\
LaTeX/BibTeX & present & arXiv source package \\
Repository & public & GitHub repository and v0.1.0 release \\
DOI archive & public & Zenodo archive with DOI \\
Artifact metadata & audited & version, author, license, and citation metadata checked \\
\bottomrule
\end{tabular}
\caption{Public artifact checklist. The archived artifact includes benchmark records, retrieval outputs, bootstrap outputs, label-audit outputs, context-utility metrics, and source files needed to regenerate the reported tables.}
\label{tab:artifact-checklist}
\end{table}

\paragraph{Artifact Availability.}
The replication package is archived on Zenodo under \href{https://doi.org/10.5281/zenodo.20176641}{doi:10.5281/zenodo.20176641}. It is mirrored in the \href{https://github.com/JJYYY-JJY/CSLibPremiseBench}{GitHub repository}, with release tag \texttt{v0.1.0}. CSLibPremiseBench evaluates source-visible CSLib proof-reference proxies, not elaborated Lean dependency traces. The reported retrieval and context-packet measurements should not be read as proof-generation or proof-repair performance.

\section{Future Proof-Repair Protocol}

The retrieval benchmark is designed to support future proof-repair experiments, but such experiments need stricter isolation than this paper can establish. A safe pilot would choose held-out tasks, remove access to original proof text and proxy-gold proof names from prompts, provide only retrieved premise context and target statements, ask a repair system to produce Lean code, and validate results by Lean compilation. Results should be reported as success/failure with full logs and negative cases, not as broad model-level proof-generation performance.

Until that protocol is implemented, this paper treats proof repair as future work. The current evidence supports benchmark construction, label-audit reporting, retrieval comparison, and context-packet analysis only.

\section{Conclusion}

CSLibPremiseBench provides a reproducible benchmark and empirical study for premise retrieval over Lean 4 computer-science theorems in CSLib. The evidence supports a careful conclusion: repository structure and candidate-policy design materially shape CSLib premise retrieval, but strong lexical baselines remain difficult to beat. CSG-Rerank gives a modest MRR gain over BM25, not a reliable win over BM25+symbol. The context-packet audit shows stronger module/family concentration, not established downstream utility. The label audit quantifies why source-visible proxy labels are useful benchmark signals but must not be described as true elaborated Lean dependencies.

The strongest contribution is therefore methodological and empirical. CSLibPremiseBench makes task construction, candidate accessibility, proxy-label limitations, statistical uncertainty, and negative results inspectable. If future work adds learned retrieval, dense embeddings, or proof repair, it should preserve this discipline through clean splits, stronger dependency evidence, and Lean-checked held-out evaluation.

\bibliographystyle{plain}
\bibliography{references}

\end{document}